\documentclass[12pt,letterpaper]{article}
\usepackage[english]{babel}
\usepackage[utf8]{inputenc}
\pdfoutput=1
\usepackage{jheppub}
\usepackage{amsfonts,amsthm,amssymb,amsfonts}
\usepackage{slashed}
\usepackage{mathrsfs}
\usepackage{hyperref}   
\usepackage{color}
\usepackage{float}
\DeclareMathOperator{\diag}{diag}
\hypersetup{unicode}

\newtheorem{proposition}{Proposition}

\newcommand{\eq}{\begin{equation}}
\newcommand{\feq}{\end{equation}}
\newcommand{\eqNN}{\begin{equation*}}
\newcommand{\feqNN}{\end{equation*}}
\font\mybb=msbm10 at 12pt
\def\bb#1{\hbox{\mybb#1}}

\def\bR {\bb{R}}

\title{Black holes in Sol minore}

\author{Federico Faedo$^{1,2}$,}
\author{Daniele Angelo Farotti$^1$}
\author{and Silke Klemm$^{1,2}$}

\affiliation{$^1$ Dipartimento di Fisica, Universit\`a di Milano, \\
Via Celoria 16, I-20133 Milano}
\affiliation{$^2$ INFN, Sezione di Milano, \\
Via Celoria 16, I-20133 Milano}

\emailAdd{federico.faedo@unimi.it}
\emailAdd{danieleangelo.farotti@studenti.unimi.it}
\emailAdd{silke.klemm@mi.infn.it}
\preprint{IFUM-1078-FT}

\abstract{We consider black holes in five-dimensional $N=2$ $\text{U}(1)$-gauged supergravity coupled
to vector multiplets, with horizons that are homogeneous but not isotropic. We write down the
equations of motion for electric and magnetic ans\"atze, and solve them explicitely for the case
of pure gauged supergravity with magnetic $\text{U}(1)$ field strength and Sol horizon.
The thermodynamics of the resulting solution, which exhibits 
anisotropic scaling, is discussed. If the horizon is compactified, the geometry approaches
asymptotically a torus bundle over AdS$_3$.
Furthermore, we prove a no-go theorem that states the nonexistence
of supersymmetric, static, Sol-invariant, electrically or magnetically charged solutions with spatial
cross-sections modelled on solve\-geometry. Finally, we study the attractor mechanism for extremal static
non-BPS black holes with nil- or solvegeometry horizons. It turns out that there are no such attractors
for purely electric field strengths, while in the magnetic case there are attractor geometries, where the
values of the scalar fields on the horizon are computed by extremization of an effective potential
$V_{\text{eff}}$, which contains the charges as well as the scalar potential of the gauged supergravity
theory. The entropy density of the extremal black hole is then given by the value of $V_{\text{eff}}$
in the extremum.
}


\begin{document}

\maketitle
\flushbottom

\section{Introduction and summary of results}

In the seventies of the last century Hawking proved his famous theorem \cite{Hawking:1971vc,
Hawking:1973uf} on the topology of black holes, which asserts that event horizon cross sections
of 4-dimensional asymptotically flat stationary black holes obeying the dominant energy condition are 
topologically $\text{S}^2$. This result extends to outer apparent horizons in black hole spacetimes that
are not necessarily stationary \cite{Hawking:72}. Such restrictive uniqueness theorems do not hold in
higher dimensions, the most famous counterexample being the black ring of Emparan and
Reall \cite{Emparan:2001wn}, with horizon topology $\text{S}^2\times\text{S}^1$.
Nevertheless, Galloway and Schoen \cite{Galloway:2005mf} were able to show that, in arbitrary
dimension, cross sections of the event horizon (in the stationary case) and outer apparent horizons
(in the general case) are of positive Yamabe type, i.e., admit metrics of positive scalar curvature.

Instead of increasing the number of dimensions, one can relax some of the assumptions that go
into Hawking's theorem in order to have black holes with nonspherical topology. One such possibility
is to add a negative cosmological constant $\Lambda$. Interpreting the term $-\Lambda g_{\mu\nu}$
as $8\pi G$ times the energy-momentum tensor $T_{\mu\nu}$, one has obviously that
$-T_{\mu\nu}\xi^\nu$ is past-pointing for every future-pointing causal vector $\xi^\nu$, and thus
a violation of the dominant energy condition. Moreover, since for $\Lambda<0$ the solutions generically
asymptote to anti-de~Sitter (AdS) spacetime, also asymptotic flatness does not hold anymore.
In this case, the horizon of a black hole can indeed be a compact Riemann surface $\Sigma_g$ of any
genus $g$ \cite{Lemos:1994xp,Mann:1996gj,Vanzo:1997gw,Cai:1996eg}.
It should be noted that, unless $g=0$, these spacetimes are asymptotically only locally
AdS; their global structure is different. This is in contrast to the black rings in five dimensions,
which are asymptotically Minkowski, in spite of their nontrivial horizon topology.
Notice in addition that the solutions of \cite{Lemos:1994xp,Mann:1996gj,Vanzo:1997gw,Cai:1996eg}
do not exhaust the spectrum of black holes in $\text{AdS}_4$,
since one can also have horizons that are noncompact manifolds with yet finite area (and thus 
finite entropy), topologically spheres with two punctures
\cite{Gnecchi:2013mja,Klemm:2014rda}\footnote{These solutions can be generalized to
$D>4$ \cite{Hennigar:2014cfa}.}.

In this paper, we will allow for both of the possibilities described above, i.e., we shall consider the
case $D=5$ and include a negative cosmological constant. More generally, our model contains
scalar fields with a potential that admits AdS$_5$ vacua. A class of uncharged black holes in
Einstein-Lambda gravity was obtained by Birmingham in \cite{Birmingham:1998nr} for arbitrary
dimension $D$. These solutions have the property that the horizon is a $(D-2)$-dimensional Einstein manifold of positive, zero, or negative curvature. In our case, $D=5$, and three-dimensional
Einstein spaces have necessarily constant curvature, i.e., are homogeneous and isotropic.
Similar to what is done in Bianchi cosmology, one can try to relax these conditions by dropping
the isotropy assumption. The horizon is then a homogeneous manifold, and belongs thus to the nine
`Bianchi cosmologies', which are in correspondence with the eight Thurston model geometries,
cf.~appendix \ref{app:homogeneous-spaces} for details. For two of these cases, namely Nil and Sol,
the corresponding black holes in five-dimensional gravity with negative cosmological constant
were constructed in \cite{Cadeau:2000tj} for the first time. Asymptotically, these solutions are neither
flat nor AdS, but exhibit anisotropic scaling.

Here we go one step further with respect to \cite{Cadeau:2000tj} and add also charge.
Some attempts in this direction include \cite{Arias:2017yqj}, where an intrinsically dyonic
black hole with Sol horizon in Einstein-Maxwell-AdS gravity was found\footnote{Since the authors 
of \cite{Arias:2017yqj} do not include a Chern-Simons term, their solution does not solve the equations
of motion of pure gauged supergravity.} and \cite{Bravo-Gaete:2017nkp}, which considers different
models that are not directly related to gauged supergravity theories.
There are various reasons for the addition of charge. First of all, charged black holes generically
have an extremal limit, and a subclass of these zero-temperature solutions might preserve
some fraction of supersymmetry, which is instrumental in holographic computations of the number
of microstates. Moreover, in the extremal limit we expect to find an attractor
mechanism \cite{Ferrara:1995ih,Strominger:1996kf,Ferrara:1996dd,Ferrara:1996um,Ferrara:1997tw},
according to which the horizon values of the scalar fields in the theory are determined by the 
electromagnetic charges alone, and do not depend on the asymptotic values of the moduli.
In our case, the corresponding attractor geometry would be $\text{AdS}_2\times M$, where $M$
denotes a three-dimensional homogeneous manifold\footnote{Note in this context that
\cite{Iizuka:2012iv} considers near-horizon geometries with various homogeneous horizons in
theories containing massive vector fields, while \cite{Kachru:2013voa} constructs solutions
interpolating between some Bianchi cosmologies and Lifshitz geometries or
$\text{AdS}_2\times\text{S}^3$. The authors of \cite{Kachru:2013voa} do not show that these
solutions can be obtained from some particular theories, but prove that the corresponding
energy-momentum tensor satisfies the weak energy condition.}.
These issues will be addressed in the following.

We start in section \ref{sec:gauged-sugra} by setting up the gauged supergravity model that will
be considered throughout the paper. In \ref{sec:eom+ans} we write down the
equations of motion for electric and magnetic ans\"atze. These are then solved explicitely in
sec.~\ref{sec:black-string} for the case of pure gauged supergravity with magnetic $\text{U}(1)$ field 
strength and Sol horizon. Moreover, the thermodynamics of the resulting solution, which exhibits 
anisotropic scaling, is discussed. If the horizon is compactified, the geometry approaches
asymptotically a torus bundle over AdS$_3$. Sec.~\ref{sec:exist-BPS} is dedicated to the proof of a
no-go theorem that states the nonexistence
of supersymmetric, static, Sol-invariant, electrically or magnetically charged solutions with spatial
cross-sections modelled on solvegeometry. Finally, in \ref{sec:attractor} we study the attractor
mechanism for extremal static
non-BPS black holes with nil- or solvegeometry horizons. It turns out that there are no such attractors
for purely electric field strengths, while in the magnetic case there are attractor geometries, where the
values of the scalar fields on the horizon are computed by extremization of an effective potential
$V_{\text{eff}}$, which contains the charges as well as the scalar potential of the gauged supergravity
theory. The entropy density of the extremal black hole is then given by the value of $V_{\text{eff}}$
in the extremum.

\section{$N=2$, $D=5$ $\text{U}(1)$-gauged supergravity}
\label{sec:gauged-sugra}

We consider $N=2$, $D=5$ $\text{U}(1)$-gauged supergravity coupled to $n$ abelian vector multiplets, 
whose bosonic field content includes the f\"unfbein $e^a_\mu$, the vectors $A^I_\mu$ with
$I=0,\ldots,n$ and the real scalars $\phi^i$, where $i=1,\ldots,n$. The gauging of the $\text{U}(1)$ 
subgroup of the $\text{SU}(2)$ R-symmetry is achieved through the vector field $A_\mu=V_I A^I_\mu$
with coupling constant $g$, where the $V_I$ are constant parameters. In order to preserve supersymmetry 
the introduction of a scalar potential is required. The bosonic part of the Lagrangian is given
by \cite{Gunaydin:1984ak}
\eq
\label{lagrangian}
e^{-1} \mathscr{L} = \frac{R}{2} - \frac12 \mathcal{G}_{ij} \partial_\mu \phi^i \partial^\mu \phi^j - \frac14 G_{IJ} F^I_{\mu\nu} F^{J\mu\nu} + \frac{e^{-1}}{48} C_{IJK}\varepsilon^{\mu\nu\rho\sigma\tau} F^I_{\mu\nu} F^J_{\rho\sigma} A^K_\tau - g^2 U \,,
\feq
where $F^I_{\mu\nu}$ are the abelian field strength tensors. The scalar potential $U$ reads
\eq
\label{scalarpot}
U = V_I V_J \bigg(\frac92 \mathcal{G}^{ij} \partial_i h^I \partial_j h^J - 6h^I h^J\bigg) \,,
\feq
where $\mathcal{G}^{ij}$ is the inverse of the target space metric $\mathcal{G}_{ij}$, $\partial_i$ denotes
the partial derivative with respect to $\phi^i$ and the functions $h^I=h^I(\phi^i)$ satisfy the condition
\eq
\label{hypersurface}
\mathcal{V} := \frac16 C_{IJK} h^I h^J h^K = 1 \,,
\feq
with $C_{IJK}$ a fully symmetric, constant and real tensor. The kinetic matrices $\mathcal{G}_{ij}$ and
$G_{IJ}$ are given by
\eq
G_{IJ} = -\frac12 \frac{\partial}{\partial{h^I}} \frac{\partial}{\partial{h^J}} \log\mathcal{V}\big|_{\mathcal{V}=1} \,,  \qquad  \mathcal{G}_{ij} = \partial_i h^I \partial_j h^J G_{IJ}\big|_{\mathcal{V}=1} \,.
\feq
The Einstein-, Maxwell-Chern-Simons- and scalar field equations following from \eqref{lagrangian}
are respectively
\eq
\label{einstein}
R_{\mu\nu} = \mathcal{G}_{ij} \partial_\mu \phi^i \partial_\nu \phi^j + G_{IJ}\bigl( F^{I\,\rho}_\mu F^J_{\nu\rho} - \frac16 g_{\mu\nu} F^I_{\rho\sigma} F^{J\rho\sigma} \bigr) + \frac23 g^2 U g_{\mu\nu} \,,
\feq
\eq
\label{maxwell}
\nabla_\lambda \bigl(G_{IJ} F^{J\lambda\tau}\bigr) + \frac{e^{-1}}{16} C_{IJK}\varepsilon^{\mu\nu\rho\sigma\tau} F^J_{\mu\nu} F^K_{\rho\sigma} = 0 \,,
\feq
\eq
\label{scalars}	
\nabla_\mu\bigl(\mathcal{G}_{ij} \partial^\mu \phi^j \bigr) - \frac12 \partial_i \mathcal{G}_{kj}\partial_\mu \phi^k \partial^\mu \phi^j - \frac14 \partial_i G_{IJ} F^I_{\mu\nu} F^{J\mu\nu} - g^2 \partial_i U = 0 \,.
\feq

\section{Equations of motion for electric and magnetic ans\"atze}
\label{sec:eom+ans}

In order to solve the equations of motion \eqref{einstein}-\eqref{scalars} we use an ansatz inspired
by \cite{Cadeau:2000tj}, with homogeneous sections $\Sigma_{t,r}$ of constant $t$ and $r$. Without loss
of generality, we take the line element to be
\eq
\label{ansatz-metric}
ds^2 = -V(r) dt^2 + \frac{dr^2}{V(r)} + \sum_{A=1}^3 e^{2T_A(r)} (\theta^A)^2 \,,
\feq
where the induced metric on $\Sigma_{t,r}$ is written in terms of $G$-invariant 1-forms $\theta^A$, which satisfy
\eq
d\theta^A = \frac12 C^A_{\ BC} \theta^B \wedge \theta^C \,,
\feq
with $C^A_{\ BC}$ the structure constants of the Lie algebra of the isometry group $G$. A list of all the possible isometry groups with related structure constants and invariant 1-forms
can be found in \cite{Ryan:1975jw}, while in appendix \ref{app:homogeneous-spaces} we present
those for solve- and nilgeometries along with a brief discussion of homogeneous manifolds. Henceforth we shall restrict our discussion to class A Bianchi models, which contain the most exotic cases, such as solve- and nilgeometry
(cf.~table~\ref{table:bianchi-thurston} in appendix \ref{app:homogeneous-spaces}).\\
The scalar fields are assumed to depend on the radial coordinate only,
\eq
\phi^i = \phi^i(r)\,.
\feq

\subsection{Electric ansatz}
For a purely electric ansatz  the vector fields are given by
\eq
\label{electric}
A^I = A^I_t(r) dt \,,
\feq
and the Maxwell equations \eqref{maxwell} imply
\eq
F^I_{rt} = \partial_r A^I_t = e^{-\sum_A T_A} G^{IJ} q_J \,,
\label{fieldel}
\feq
where $G^{IJ}$ denotes the inverse of $G_{IJ}$, and the constants $q_I$ represent essentially the electric
charge densities.\\
Using \eqref{fieldel} and the Bianchi class A condition \eqref{classA}, the Einstein equations
\eqref{einstein} boil down to
\eq
\label{einstein-el}
\begin{gathered}
\frac{V''}{2} + \frac{V'}{2} \sum_A T_A' = \frac23 e^{-2\sum_A T_A} G^{IJ} q_I q_J - \frac23 g^2 U \,, \\
\sum_A T_A'' + \sum_A (T_A')^2 = -\mathcal{G}_{ij} {\phi^i}' {\phi^j}' \,, \qquad
\sum_B C^B_{\ AB} T_B' = 0 \,, \\
-V' T_A' - V T_A'' - V T_A' \sum_B T_B' + \mathcal{J}_A = \frac13 e^{-2\sum_B T_B}G^{IJ} q_I q_J + \frac23 g^2U \,,
\end{gathered}
\feq
where we defined
\eq
\mathcal{J}_A := \sum_{B,C} \left[ -\frac12 D^B_{\ AC} \bigl(D^C_{\ AB} + D^B_{\ AC}\bigr) +\frac14 \bigl(D^A_{\ BC}\bigr)^2 \right] \,,
\feq
with
\eq
D^A_{\ BC} := e^{T_A-T_B-T_C} C^A_{\ BC} \,.
\feq
The third equation in \eqref{einstein-el} is a constraint, which is trivially satisfied for all the class A Bianchi
cosmologies except for solvegeometry; in this case it reduces to
\eq
\label{vincolo}
T_1' = T_2' \,.
\feq
Finally, using \eqref{fieldel}, the equations \eqref{scalars} for the scalars become
\eq
\begin{split}
& V\mathcal{G}_{ij} {\phi^j}'\sum_A T'_A + V \frac{d\mathcal{G}_{ij}}{dr} {\phi^j}' + V' \mathcal{G}_{ij} {\phi^j}' + V \mathcal{G}_{ij} {\phi^j}'' - \frac12 V \partial_i \mathcal{G}_{kj} {\phi^k}'
{\phi^j}' \\
& \qquad -\frac12 e^{-2\sum_A T_A}\partial_i G^{IJ} q_I q_J - g^2 \partial_i U = 0 \,.
\end{split}
\feq

\subsection{Magnetic ansatz}

In the magnetically charged case we take for the field strength
\eq
F^I = p^I\theta^1\wedge\theta^2\,,
\feq
where the $p^I$ are magnetic charge densities. Note that $F^I$ is closed due to the
Bianchi class A condition \eqref{classA}, so locally there exists a gauge potential $A^I$ such that
$F^I=dA^I$. In the following we shall consider the case of solvegeometry, for which
\eq
\label{magnetic}
F^I = p^I dx\wedge dy\,, \qquad A^I = p^I x dy\,.
\feq
Using \eqref{Solforms} the line element \eqref{ansatz-metric} becomes
\eq
ds^2 = -V(r) dt^2 + \frac{dr^2}{V(r)} + e^{2(T_1(r)+z)} dx^2 + e^{2(T_2(r)-z)} dy^2 + e^{2T_3(r)} dz^2 \,.
\label{metric:sol}
\feq
The Maxwell equations \eqref{maxwell} are automatically satisfied by \eqref{magnetic} and
\eqref{metric:sol}, while the nontrivial Einstein equations \eqref{einstein} read
\eq
\label{einstein-magn}
\begin{gathered}
\frac{V''}{2} + \frac{V'}{2} \bigl(2T_1'+T_3'\bigr) = \frac13 e^{-4T_1} G_{IJ} p^I p^J - \frac23 g^2 U \,, \\
2T_1'' + T_3'' + 2(T_1')^2 + (T_3')^2 = -\mathcal{G}_{ij} {\phi^i}' {\phi^j}' \,, \\
-V' T_1' - V \bigl(T_1'' + T_1'(2T_1'+T_3')\bigr) = \frac23 e^{-4T_1} G_{IJ} p^I p^J + \frac23 g^2 U \,, \\
-V' T_3' - V \bigl(T_3'' + T_3'(2T_1'+T_3')\bigr) - 2e^{-2T_3} = -\frac13 e^{-4T_1} G_{IJ} p^I p^J +\frac23 g^2 U \,,
\end{gathered}
\feq
where we have used the condition $T_1'=T_2'$ and the freedom to rescale $y$ in order to set $T_1=T_2$. The scalar field equations \eqref{scalars} become
\eq
\label{scalars-magn}
\begin{split}
& V\mathcal{G}_{ij}{\phi^j}'\sum_A T_A' + V \frac{d\mathcal{G}_{ij}}{dr} {\phi^j}' + V' \mathcal{G}_{ij} {\phi^j}' + V \mathcal{G}_{ij} {\phi^j}'' - \frac12 V \partial_{i} \mathcal{G}_{kj} {\phi^k}' {\phi^j}' \\
& \qquad - \frac12 \partial_i G_{IJ} e^{-4T_1} p^I p^J - g^2 \partial_i U = 0 \,.
\end{split}
\feq

\section{Magnetic black hole in pure gauged supergravity}
\label{sec:black-string}

In order to study the above equations in a simplified setting, we restrict our attention to
pure gauged supergravity, i.e., the theory \eqref{lagrangian} without vector multiplets ($n=0$).
For a purely electric or magnetic configuration, the Chern-Simons term can be consistently
truncated, and \eqref{lagrangian} boils down to
\eq
e^{-1} \mathscr{L} = \frac{R}{2} - \frac14 F_{\mu\nu} F^{\mu\nu} - \Lambda \,,
\feq
where $\Lambda=-6g^2<0$, $F_{\mu\nu}=F^0_{\mu\nu}$ and we fixed $C_{000}$ in \eqref{hypersurface} 
and $V_0$ in \eqref{scalarpot} such that $G_{00}=1$ and $U=-6$.\\
The field strength \eqref{magnetic} becomes simply $F_{xy}=p$, while the Einstein equations
\eqref{einstein-magn} reduce to 
\eq
\begin{gathered}
\frac{V''}{2} + \frac{V'}{2} \bigl(2T_1'+T_3'\bigr) = \frac13 e^{-4T_1} p^2 - \frac23 \Lambda \,, \\
2T_1'' + T_3'' + 2(T_1')^2 + (T_3')^2 = 0 \,, \\
-V' T_1' - V \bigl(T_1'' + T_1'(2T_1'+T_3')\bigr) = \frac23 e^{-4T_1} p^2 + \frac23 \Lambda \,, \\
-V' T_3' - V \bigl(T_3'' + T_3'(2T_1'+T_3')\bigr) - 2e^{-2T_3} = -\frac13 e^{-4T_1} p^2 + \frac23
\Lambda\,.
\end{gathered}
\label{puremagn}
\feq
One easily checks that in the uncharged case $p=0$ the above equations are satisfied by the solvegeometry 
solution constructed in \cite{Cadeau:2000tj}.\\
\eqref{puremagn} can be easily solved by taking $T_1$ to be constant\footnote{Note that this is not the
case for the solution of \cite{Cadeau:2000tj}.}.
With this assumption, a particular black hole solution is given by
\eq
ds^2 = -V(r) dt^2 + \frac{dr^2}{V(r)} + \sqrt{\frac{p^2}{-\Lambda}} \bigl(e^{2z} dx^2 + e^{-2z}dy^2\bigr) + \frac{r^2}{A} dz^2 \,,
\label{blackstring}
\feq
\eq
F = p dx\wedge dy\,,
\label{fieldstrength}
\feq
with
\eq
V(r) = -\frac{\Lambda}{2} r^2 - 2A\ln\left(\frac{r}{B}\right)\,,
\label{V}
\feq
where $A$ and $B$ are two positive integration constants. It is worth noting that this solution is singular
in the limit $p\to0$, and it is thus disconnected from the one in \cite{Cadeau:2000tj}.
The metric \eqref{blackstring} and field strength \eqref{fieldstrength} are invariant under the scale 
transformations
\eq
t\to t/\nu\,, \qquad r\to \nu r\,, \qquad z\to z + \ln\alpha\,, \qquad x\to
\lambda x\,, \qquad y\to \pm\lambda\alpha^2 y\,, \label{scaling-symm}
\feq
accompanied by
\eq
p\to\pm\frac p{\lambda^2\alpha^2}\,, \qquad A\to\nu^2 A\,, \qquad B\to\nu B\,.
\feq
This can be used to set e.g.~$p=B=1/g$ without loss of generality. $B$ and the magnetic charge
density $p$ are thus not true parameters of the solution, which is specified completely by
choosing $A$. Notice that the scaling symmetries with $\nu=1$, $\lambda=1/\alpha$ belong to
the Lie group Sol. If the horizon is compactified (cf.~\cite{Cadeau:2000tj} for details on the
compactification procedure), the transformations in \eqref{scaling-symm} involving $\alpha$ and
$\lambda$ are broken down to a discrete subgroup ($\alpha=\lambda^{-1}=e^{na}$, where $a$
is the constant appearing in (II.22) of \cite{Cadeau:2000tj} and $n\in\mathbb{Z}$), which does no more
allow to scale $p$ to any value. In this case, $p$ can become actually a genuine parameter of
the black hole.\\
\eqref{blackstring} exhibits anisotropic scaling. If the horizon is compactified, the geometry approaches
asymptotically for $r\to\infty$ a torus bundle over AdS$_3$.
In $r=0$ there is a curvature singularity, since the Kretschmann scalar behaves as $R^{\mu\nu\rho\sigma} 
R_{\mu\nu\rho\sigma}\sim (\ln r)/r^4$ for $r\to 0$. Horizons are determined by the roots of the
function $V(r)$, which diverges both for $r\to0$ and $r\to+\infty$ and has a unique minimum in
\eq
r = r_{\text{min}} =\sqrt{\frac{2A}{-\Lambda}}\,.
\feq
If $V(r_{\text{min}})>0$ the solution represents a naked singularity. For $V(r_{\text{min}})=0$, i.e.,
$A=3e$, we have an extremal black hole, while for $V(r_{\text{min}})<0$ ($A>3e$) there is an inner
and an outer horizon and the solution is nonextremal.\\
Requiring the absence of conical singularities in the Euclidean section gives the Hawking temperature
\eq
T = \frac{-\Lambda r_{\text h}^2 - 2A}{4\pi r_{\text h}}\,,
\feq
where $r_{\text h}$ denotes the radial coordinate of the horizon.
The entropy density can be computed by means of the Bekenstein-Hawking formula and is given by
\eq
s = \frac{S}{V_{\text{solve}}} = \frac{(\ln(g r_{\text h}))^{1/2}}{12 g^3}\,,
\feq
where we set Newton's constant $G=1$, and
$V_{\text{solve}}$ is the volume of the compactified manifold modelled on solvegeometry.\\
The standard Komar integral for the mass goes like $\Lambda r^2$ for large $r$ and thus diverges for $r\to+\infty$ due to the presence of the vacuum energy, as was to be expected. Moreover, there is
no obvious background to subtract, and the conditions for the applicability of the Ashtekar-Magnon-Das 
formalism \cite{Ashtekar:1984zz,Ashtekar:1999jx} are not satisfied. In spite of these difficulties, we
can associate a mass to the black hole \eqref{blackstring} by simply integrating the first law. Since
$p$ is not a dynamical parameter of the solution, we do not expect a term containing the variation
of the magnetic charge in the first law. The mass density $m$ satisfies thus
\eq
dm = Tds\,,
\feq
which gives (up to an integration constant, that can be fixed by requiring e.g.~the extremal
solution to have zero energy)
\eq
m = \frac{r_{\text h}}{16\pi g(\ln(g r_{\text h}))^{1/2}}\,.
\feq
To close this section, we remark that a generalization of the solution \eqref{blackstring}, 
\eqref{fieldstrength} as well as the one of \cite{Cadeau:2000tj} to the stu model of $N=2$ gauged 
supergravity, together with a numerical analysis of the equations of motion \eqref{puremagn},
is currently under investigation.

\section{Existence of static, Sol-invariant BPS solutions}
\label{sec:exist-BPS}

A simpler method to construct solutions to a given supergravity theory is based on solving the
Killing spinor equations. These are of first order, and are generically much easier to solve than
the full second order equations of motion. At least in the case where the Killing vector constructed
as a bilinear from the Killing spinor is timelike, the latter are implied by the Killing spinor
equations \cite{Gutowski:2004yv}.\\
The supersymmetry variations for the gravitino $\psi_\mu$ and the gauginos $\lambda_i$ in a bosonic 
background are given by (see e.g.~\cite{Klemm:2000nj})
\begin{align}
\label{gravitino}
\delta\psi_\mu & = \left[ \mathcal{D}_\mu + \frac{i}{8} h_I \big(\Gamma_\mu^{\ \nu\rho} - 4\delta_\mu^{\ \nu} \Gamma^\rho\big) F^I_{\nu\rho} + \frac{g}{2} \Gamma_\mu h^I V_I - \frac{3i}{2}g V_I A^I_\mu \right] \epsilon \,, \\
\label{gauginos}
\delta\lambda_i & = \left[ \frac38 \Gamma^{\mu\nu} F_{\mu\nu}^I \partial_i h_I - \frac{i}{2} \mathcal{G}_{ij} \Gamma^\mu \partial_\mu \phi^j + \frac{3i}{2} g V_I \partial_i h^I \right] \epsilon \,, 
\end{align}
where $\epsilon$ is the supersymmetry parameter, $h_I=\frac16 C_{IJK} h^J h^K$ and $\mathcal{D}_\mu$
denotes the Lorentz-covariant derivative\footnote{Our conventions are $\mathcal{D}_\mu=\partial_\mu
+ \frac14\omega_\mu^{\ ab}\Gamma_{ab}$, $\{\Gamma^a,\Gamma^b\}=2\eta^{ab}$,
$\Gamma^{a_1 a_2 \ldots a_n}=\Gamma^{[a_1} \Gamma^{a_2} \ldots \Gamma^{a_n]}$, where we
antisymmetrize with unit weight.}.
The vanishing of the gravitino supersymmetry transformations~\eqref{gravitino} leads to the Killing
spinor equations, whose integrability conditions imply a set of constraints for the metric and the matter 
fields. Given $\delta\psi_\mu\equiv\hat{\mathcal{D}}_\mu \epsilon=0$, the first integrability conditions read
\eq
\label{integrability}
\hat{\mathcal{R}}_{\mu\nu} \epsilon\equiv\bigl[ \hat{\mathcal{D}}_\mu, \hat{\mathcal{D}}_\nu \bigr] \epsilon = 0\,,
\feq
which is a set of algebraic equations that admit a nontrivial solution $\epsilon$ iff 
$\det(\hat{\mathcal{R}}_{\mu\nu})=0$.\\
In what follows we shall specify to solvegeometry with electric or magnetic ansatz. For the metric 
\eqref{metric:sol} the tetrad can be chosen as
\eq
e^0_t = \sqrt{V} \,, \qquad  e^1_x = e^{T_1+z} \,, \qquad  e^2_y = e^{T_2-z} \,, \qquad  e^3_z =
e^{T_3} \,, \qquad  e^4_r = \frac1{\sqrt{V}} \,.
\feq

\subsection{Electric ansatz}

In the case of solvegeometry and electric ansatz, the vanishing of the gravitino variations
\eqref{gravitino} leads to
\eq
\label{KSE:el}
\begin{split}
&\left[ \partial_t + \frac{V'}{4} \, \Gamma_{04} + \frac{i}{2} h_I F^I_{rt} \sqrt{V} \, \Gamma_4 + \frac{g}{2} V_I h^I \sqrt{V} \, \Gamma_0 - i\frac{3g}{2} V_I A^I_t \right]\epsilon = 0\,, \\
&\left[ \partial_r + \frac{i}{2} h_I F^I_{rt} \frac{1}{\sqrt{V}} \, \Gamma_0 + \frac{g}{2} V_I h^I \frac{1}{\sqrt{V}} \, \Gamma_4 \right]\epsilon = 0\,, \\
&\left[ \partial_x + e^{T_1+z} \left( \frac12 \sqrt{V} T_1' \, \Gamma_{14} + \frac12 e^{-T_3} \, \Gamma_{13} - \frac{i}{4} h_I F^I_{rt} \, \Gamma_{014} + \frac{g}{2} V_I h^I \, \Gamma_1 \right) \right]\epsilon = 0\,, \\
&\left[ \partial_y + e^{T_2-z} \left( \frac12 \sqrt{V} T_2' \, \Gamma_{24} - \frac12 e^{-T_3} \, \Gamma_{23} - \frac{i}{4} h_I F^I_{rt} \, \Gamma_{024} + \frac{g}{2} V_I h^I \, \Gamma_2 \right) \right]\epsilon = 0\,, \\
&\left[ \partial_z + e^{T_3} \left( \frac12 \sqrt{V} T_3' \, \Gamma_{34} - \frac{i}{4} h_I F^I_{rt} \, \Gamma_{034} + \frac{g}{2} V_I h^I \, \Gamma_3 \right) \right]\epsilon = 0\,.
\end{split}
\feq
The integrability conditions \eqref{integrability} with $(\mu,\nu)$ equal to $(t,x)$, $(t,y)$ and $(t,z)$ are, respectively,
\eq
\label{el:integrability-time}
\begin{gathered}
\left[ \frac12 V' T_1' - g^2 (V_I h^I)^2 + i\sqrt{V} T_1' h_I F^I_{rt} \, \Gamma_0 + ig V_I h^I h_J F^J_{rt}\, \Gamma_{04} \right] \epsilon = 0 \,, \\
\left[ \frac12 V' T_2' - g^2 (V_I h^I)^2 + i\sqrt{V} T_2' h_I F^I_{rt} \, \Gamma_0 + ig V_I h^I h_J F^J_{rt}\, \Gamma_{04} \right] \epsilon = 0 \,, \\
\left[ \frac12 V' T_3' - g^2 (V_I h^I)^2 + i\sqrt{V} T_3' h_I F^I_{rt} \, \Gamma_0 + ig V_I h^I h_J F^J_{rt}\, \Gamma_{04} \right] \epsilon = 0 \,,
\end{gathered}
\feq
while for $(x,y)$, $(x,z)$ and $(y,z)$ we have
\begin{align}
\label{el:integrability-space}
& \begin{split}
  & \biggl[ V T_1' T_2' - g^2 (V_I h^I)^2 + \frac14 (h_I F^I_{rt})^2 - e^{-2T_3} - \frac i2\sqrt{V} (T_1'+T_2')
h_I F^I_{rt}\, \Gamma_0 \\
  & \qquad - ig V_I h^I h_J F^J_{rt}\, \Gamma_{04} \biggr] \epsilon = 0 \,,
\end{split} \notag \\
\displaybreak[0]
& \begin{split}
  & \biggl[ V T_1' T_3' - g^2 (V_I h^I)^2 + \frac14 (h_I F^I_{rt})^2 + e^{-2T_3} - \frac i2\sqrt{V} (T_1'+T_3')
h_I F^I_{rt}\, \Gamma_0 \\
  & \qquad - ig V_I h^I h_J F^J_{rt}\, \Gamma_{04} + \sqrt{V} (T_1'-T_3') e^{-T_3} \, \Gamma_{34} \biggr] 
\epsilon = 0 \,,
\end{split} \\
\displaybreak[0]
& \begin{split}
  & \biggl[ V T_2' T_3' - g^2 (V_I h^I)^2 + \frac14 (h_I F^I_{rt})^2 + e^{-2T_3} - \frac i2\sqrt{V} (T_2'+T_3')
h_I F^I_{rt}\, \Gamma_0 \\
  & \qquad - ig V_I h^I h_J F^J_{rt}\, \Gamma_{04} - \sqrt{V} (T_2'-T_3') e^{-T_3} \, \Gamma_{34} \biggr] \epsilon = 0 \,.
\end{split} \notag 
\end{align}
The difference of eqns.~\eqref{el:integrability-time} taken in (all the three possible) pairs leads to
\eq
\begin{gathered}
\label{el:integrability-time-combi}
(T_1'-T_2') \left[ \frac12 V' + i\sqrt{V} h_I F^I_{rt} \, \Gamma_0 \right] \epsilon = 0 \,, \\
(T_1'-T_3') \left[ \frac12 V' + i\sqrt{V} h_I F^I_{rt} \, \Gamma_0 \right] \epsilon = 0 \,, \\
(T_2'-T_3') \left[ \frac12 V' + i\sqrt{V} h_I F^I_{rt} \, \Gamma_0 \right] \epsilon = 0 \,,
\end{gathered}
\feq
whereas $(x,y)-(x,z)$ and $(x,y)-(y,z)$ read
\eq
\begin{gathered}
\label{el:integrability-space-combi}
\left[ V T_1'(T_2'-T_3') - 2e^{-2T_3} - \frac{i}{2} \sqrt{V} (T_2'-T_3') h_I F^I_{rt}\, \Gamma_0 - \sqrt{V} (T_1'-T_3') e^{-T_3} \, \Gamma_{34} \right] \epsilon = 0 \,, \\
\left[ V T_2'(T_1'-T_3') - 2e^{-2T_3} - \frac{i}{2} \sqrt{V} (T_1'-T_3') h_I F^I_{rt}\, \Gamma_0 + \sqrt{V} (T_2'-T_3') e^{-T_3} \, \Gamma_{34} \right] \epsilon = 0 \,.
\end{gathered}
\feq
We can distinguish between two different cases in which \eqref{el:integrability-time-combi} hold.
\begin{itemize}
\item{Case A}
\eq
T_1' = T_2' = T_3' \,.
\feq
In this case \eqref{el:integrability-space-combi} leads directly to the trivial solution $\epsilon = 0$.
\item{Case B}
\eq
\label{el:integrability:B}
\left[ \frac12 V' + i\sqrt{V} h_I F^I_{rt} \, \Gamma_0 \right] \epsilon = 0 \,.
\feq
Writing this condition schematically as $\mathcal{M}\epsilon=0$, a necessary condition to have nontrivial
solutions is $\det\mathcal{M}=0$, and thus
\eq
\frac12 V' = \pm \sqrt{V} h_I F^I_{rt} \,,
\feq
which, once plugged back into \eqref{el:integrability:B} gives the projection
\eq
\Gamma_0 \epsilon = \pm i\epsilon \,. \label{proj-Gamma0}
\feq
Using \eqref{proj-Gamma0} in \eqref{el:integrability-space-combi}, we get
\begin{displaymath}
\begin{gathered}
\left[ V T_1'(T_2'-T_3') - 2e^{-2T_3} \pm \frac12 \sqrt{V} (T_2'-T_3') h_I F^I_{rt} - \sqrt{V} (T_1'-T_3') e^{-T_3} \, \Gamma_{34} \right] \epsilon = 0 \,, \\
\left[ V T_2'(T_1'-T_3') - 2e^{-2T_3} \pm \frac12 \sqrt{V} (T_1'-T_3') h_I F^I_{rt} + \sqrt{V} (T_2'-T_3') e^{-T_3} \, \Gamma_{34} \right] \epsilon = 0 \,.
\end{gathered}
\end{displaymath}
To have nontrivial solutions, the determinants of the two coefficient matrices in these linear systems
must vanish, leading to $T_1' = T_3'$ and $T_2' = T_3'$, which brings us back to case A.
\end{itemize}
We can thus state the following
\begin{proposition}
\label{theo:no-go-el}
There are no static, Sol-invariant solutions to the Killing spinor equations with solvegeometry spatial
cross-sections at fixed $r$ and purely electric field strengths.
\end{proposition}

\subsection{Magnetic ansatz}
\label{subsec:magBPS}

In this case, the Killing spinor equations become
\begin{eqnarray}
&&\left[ \partial_t + \frac{V'}{4} \, \Gamma_{04} + \frac{i}{4} h_I p^I \sqrt{V} e^{-T_1-T_2} \, \Gamma_{012} + \frac{g}{2} V_I h^I \sqrt{V} \, \Gamma_0 \right] \epsilon = 0\,, \nonumber \\
&&\left[ \partial_r + \frac{i}{4} h_I p^I \frac{1}{\sqrt{V}} e^{-T_1-T_2} \, \Gamma_{124} + \frac{g}{2} V_I h^I \frac{1}{\sqrt{V}} \, \Gamma_4 \right] \epsilon = 0\,, \nonumber \\
&&\left[ \partial_x + e^{T_1+z} \left( \frac12 \sqrt{V} T_1' \, \Gamma_{14} + \frac12 e^{-T_3} \, \Gamma_{13} - \frac{i}{2} h_I p^I e^{-T_1-T_2} \, \Gamma_2 + \frac{g}{2} V_I h^I \, \Gamma_1 \right) \right] \epsilon = 0\,, \nonumber \\
&&\left[ \partial_y + e^{T_2-z} \left( \frac12 \sqrt{V} T_2' \, \Gamma_{24} - \frac12 e^{-T_3} \, \Gamma_{23} + \frac{i}{2} h_I p^I e^{-T_1-T_2} \, \Gamma_1 + \frac{g}{2} V_I h^I \, \Gamma_2  \right) - i\frac{3g}{2} V_I p^I x \right] \epsilon = 0\,, \nonumber \\
&&\left[ \partial_z + e^{T_3} \left( \frac12 \sqrt{V} T_3' \, \Gamma_{34} + \frac{i}{4} h_I p^I e^{-T_1-T_2} \, \Gamma_{123} + \frac{g}{2} V_I h^I \, \Gamma_3 \right) \right] \epsilon = 0\,.
\end{eqnarray}
We have thus the following first integrability conditions:
\begin{itemize}
\item{(t,x)}
\begin{eqnarray}
&&\biggl[ \frac12 V' T_1' - g^2 (V_I h^I)^2 + \frac i2 h_I p^I\sqrt{V} T_1' e^{-T_1-T_2}\Gamma_{124} + \frac i2 h_I p^I e^{-T_1-T_2-T_3} \Gamma_{123}\nonumber \\
&&\quad + ig h_I p^I V_J h^J e^{-T_1-T_2} \Gamma_{12} \biggr] \epsilon = 0 \,, \label{magn-tx}
\end{eqnarray}
\item{(t,y)}
\begin{eqnarray}
&&\biggl[ \frac12 V' T_2' - g^2 (V_I h^I)^2 + \frac i2 h_I p^I\sqrt{V} T_2' e^{-T_1-T_2} \Gamma_{124} - \frac i2 h_I p^I e^{-T_1-T_2-T_3} \Gamma_{123}\nonumber \\
&&\quad + ig h_I p^I V_J h^J e^{-T_1-T_2} \Gamma_{12} \biggr] \epsilon = 0 \,, \label{magn-ty}
\end{eqnarray}
\item{(t,z)}
\eq
\biggl[ \frac12 V' T_3' - g^2 (V_I h^I)^2 - ig h_I p^I V_J h^J e^{-T_1-T_2} \Gamma_{12} - \frac14 (h_I p^I)^2 e^{-2(T_1+T_2)} \biggr] \epsilon = 0 \,, \label{magn-tz}
\feq
\item{(x,y)}
\begin{eqnarray}
&&\biggl[ V T_1' T_2' - g^2 (V_I h^I)^2 + (h_I p^I)^2 e^{-2(T_1+T_2)} - e^{-2T_3}\nonumber \\
&&\quad - i h_I p^I\sqrt{V} (T_1'+T_2') e^{-T_1-T_2} \Gamma_{124} - 3ig V_I p^I e^{-T_1-T_2} \Gamma_{12} \biggr] \epsilon = 0 \,, \label{magn-xy}
\end{eqnarray}
\item{(x,z)}
\begin{eqnarray}
&&\biggl[ V T_1' T_3' - g^2 (V_I h^I)^2 + e^{-2T_3} + \sqrt{V} (T_1'-T_3') e^{-T_3} \Gamma_{34} - i h_I p^I e^{-T_1-T_2-T_3} \Gamma_{123}\nonumber \\
&&\quad + \frac i2 h_I p^I\sqrt{V} T_1' e^{-T_1-T_2} \Gamma_{124} + ig h_I p^I V_J h^J e^{-T_1-T_2} \Gamma_{12} \biggr] \epsilon = 0 \,, \label{magn-xz}
\end{eqnarray}
\item{(y,z)}
\begin{eqnarray}
&&\biggl[ V T_2' T_3' - g^2 (V_I h^I)^2 + e^{-2T_3} - \sqrt{V} (T_2'-T_3') e^{-T_3} \Gamma_{34} + i h_I p^I e^{-T_1-T_2-T_3} \Gamma_{123}\nonumber \\
&&\quad + \frac i2 h_I p^I\sqrt{V} T_2' e^{-T_1-T_2} \Gamma_{124} + ig h_I p^I V_J h^J e^{-T_1-T_2} \Gamma_{12} \biggr] \epsilon = 0 \,, \label{magn-yz}
\end{eqnarray}
\item{(r,t)}
\begin{eqnarray}
&&\biggl[ \frac12 V'' - g^2 (V_I h^I)^2 - \frac14 (h_I p^I)^2 e^{-2(T_1+T_2)} - \frac i2 h_I p^I\sqrt{V} (T_1'+T_2') e^{-T_1-T_2} \Gamma_{124}\nonumber \\
&&\quad - \frac i2\partial_r (h_I p^I) \sqrt{V} e^{-T_1-T_2} \Gamma_{124} + g \partial_r (V_I h^I) \sqrt{V} \Gamma_4\nonumber \\
&&\quad - ig h_I p^I V_J h^J e^{-T_1-T_2} \Gamma_{12} \biggr] \epsilon = 0 \,, \label{magn-rt}
\end{eqnarray}
\item{(r,x)}
\begin{eqnarray}
&&\biggl[ V T_1'' + V {T_1'}^2 - i \partial_r (h_I p^I) \sqrt{V} e^{-T_1-T_2} \Gamma_{124} + g \partial_r (V_I h^I) \sqrt{V} \Gamma_4\nonumber \\
&&\quad - \frac i2 h_I p^I\sqrt{V} (T_1'-2T_2') e^{-T_1-T_2} \Gamma_{124} \biggr] \epsilon = 0 \,,
\label{magn-rx}
\end{eqnarray}
\item{(r,y)}
\begin{eqnarray}
&&\biggl[ V T_2'' + V {T_2'}^2 - i \partial_r (h_I p^I) \sqrt{V} e^{-T_1-T_2} \Gamma_{124} + g \partial_r (V_I h^I) \sqrt{V} \Gamma_4\nonumber \\
&&\quad - \frac i2 h_I p^I\sqrt{V} (T_2'-2T_1') e^{-T_1-T_2} \Gamma_{124} \biggr] \epsilon = 0 \,,
\label{magn-ry}
\end{eqnarray}
\item{(r,z)}
\begin{eqnarray}
&&\biggl[ V T_3'' + V {T_3'}^2 + \frac{i}{2} \partial_r (h_I p^I) \sqrt{V} e^{-T_1-T_2} \Gamma_{124} + g \partial_r (V_I h^I) \sqrt{V} \Gamma_4\nonumber \\
&&\quad - \frac i2 h_I p^I\sqrt{V} (T_1'+T_2') e^{-T_1-T_2} \Gamma_{124} \biggr] \epsilon = 0 \,.
\label{magn-rz}
\end{eqnarray}
\end{itemize}
From the vanishing of the gaugino variation \eqref{gauginos} one gets
\eq
\biggl[ \frac13 \mathcal{G}_{ij} \sqrt{V} \partial_r \phi^j \Gamma_4 - g \partial_i (V_I h^I) + \frac i2 
\partial_i (h_I p^I) e^{-T_1-T_2} \Gamma_{12} \biggr] \epsilon = 0 \,. \label{magn-gaugini}
\feq
The combination \eqref{magn-xz} + \eqref{magn-yz} - \eqref{magn-tx} - \eqref{magn-ty} gives
\eq
\biggl[ (T_1'+T_2') \Bigl(V T_3' - \frac12 V'\Bigr) + 2e^{-2T_3} + \sqrt{V} (T_1'-T_2') e^{-T_3} \Gamma_{34} \biggr] \epsilon = 0 \,.
\feq
The determinant of the coefficient matrix of this linear system vanishes if
\eq
\label{T1=T2}
(T_1'+T_2') \Bigl(V T_3' - \dfrac12 V'\Bigr) + 2e^{-2T_3} = 0\quad\wedge\quad
\sqrt V (T_1'-T_2') e^{-T_3} = 0\,,
\feq
which implies
\eq
T_1' \Bigl(V T_3' - \dfrac12 V'\Bigr) + e^{-2T_3} = 0\,, \qquad
T_1' = T_2'\,.
\feq
From the combination \eqref{magn-rx} - \eqref{magn-ry} + \eqref{magn-xz} - \eqref{magn-yz}
+ 2$\cdot$(\eqref{magn-tx} - \eqref{magn-ty}) we obtain
\eq
\begin{split}
& \biggl[ V (T_1''-T_2'') + (T_1'-T_2') \bigl(V(T_1'+T_2'+T_3') + V'\bigr) + \sqrt{V} (T_1'+T_2'-2T_3') e^{-T_3} \Gamma_{34} \biggr] \epsilon = 0\,. \label{eq-imply-T_3'=T_1'}
\end{split}
\feq
Using $T_1'=T_2'$, it turns out that the vanishing of the determinant associated to
\eqref{eq-imply-T_3'=T_1'} requires $T_3'=T_1'$. \eqref{magn-tx} - \eqref{magn-ty} yields
\eq
i h_I p^I e^{-T_1-T_2-T_3}\Gamma_{123}\epsilon = 0\,,
\feq
and thus
\eq
h_I p^I = 0\,. \label{central-charge}
\feq
Taking into account the above results and defining $T'\equiv T_1'=T_2'=T_3'$, the first integrability
conditions become
\begin{itemize}
\item{(t,x), (t,y), (t,z)}
\eq
\label{KSE:mag:txyz}
\biggl[ \frac12 V' T' - g^2 (V_I h^I)^2 \biggr] \epsilon = 0 \,,
\feq
\item{(x,y)}
\eq
\label{KSE:mag:xy}
\biggl[ V {T'}^2 - g^2 (V_I h^I)^2 - e^{-2T_3} - 3ig V_I p^I e^{-T_1-T_2} \Gamma_{12} \biggr]
\epsilon = 0\,,
\feq
\item{(x,z), (y,z)}
\eq
\label{KSE:mag:xz}
\biggl[ V {T'}^2 - g^2 (V_I h^I)^2 + e^{-2T_3} \biggr] \epsilon = 0 \,,
\feq
\item{(r,t)}
\eq
\label{KSE:mag:rt}
\biggl[ \frac12 V'' - g^2 (V_I h^I)^2 + g \partial_r (V_I h^I) \sqrt{V} \Gamma_4 \biggr] \epsilon = 0 \,,
\feq
\item{(r,x), (r,y), (r,z)}
\eq
\label{KSE:mag:rxyz}
\biggl[ V T'' + V {T'}^2 + g \partial_r (V_I h^I) \sqrt{V} \Gamma_4 \biggr] \epsilon = 0 \,.
\feq
\end{itemize}
\eqref{KSE:mag:xy} - \eqref{KSE:mag:xz} leads to
\eq
\biggl[ 2e^{-2T_3} + 3ig V_I p^I e^{-T_1-T_2} \Gamma_{12} \biggr] \epsilon = 0 \,, \label{Dirac}
\feq
which implies the Dirac-type quantization condition
\eq
V_I p^I = \sigma_1 \frac{2}{3g} e^{T_1+T_2-2T_3}\,,
\feq
where $\sigma_1=\pm 1$. Plugging this back into \eqref{Dirac} gives
\eq
\Gamma_{12}\epsilon = i\sigma_1\epsilon\,. \label{proj-Gamma12}
\feq
With \eqref{proj-Gamma12}, the gaugino equation \eqref{magn-gaugini} becomes
\eq
\biggl[ \frac13 \mathcal{G}_{ij} \sqrt{V} \partial_r \phi^j \Gamma_4 - g \partial_i (V_I h^I) - \sigma_1 \frac12 \partial_i (h_I p^I) e^{-T_1-T_2} \biggr] \epsilon = 0\,. \label{eq:gaugini-proj}
\feq
If the scalar fields were constant, $\partial_r\phi^j=0\;\forall j$, this would imply
\eq
g \partial_i (V_I h^I) + \sigma_1 \frac12 \partial_i (h_I p^I) e^{-T_1-T_2} = 0\,,
\feq
and thus $T_1$ and $T_2$ must be constant as well, which leads to a contradiction with the first
equation of \eqref{T1=T2}. Note that this conclusion is valid provided $\partial_i(V_I h^I)$ and
$\partial_i(h_I p^I)$ do not both vanish. In the latter case, however, using one of the very special
geometry relations, we have
\eq
0 = \mathcal{G}^{ij}\partial_i (h_I p^I)\partial_j (h_J p^J) = \frac49 G_{IJ} p^I p^J - \frac23 h_I p^I h_J p^J
= \frac49 G_{IJ} p^I p^J\,, \label{eq:contradict}
\feq
where the last step follows from \eqref{central-charge}. Since $G_{IJ}$ is positive definite,
\eqref{eq:contradict} leads to a contradiction. If $\partial_r\phi^i\neq0$ for at least one $i$, one can
multiply \eqref{eq:gaugini-proj} with $\partial_r\phi^i$ and sum over $i$ to get\footnote{Notice that
$\partial_r(h_I p^I)=0$.}
\eq
\biggl[ \frac13 \mathcal{G}_{ij} \sqrt{V} \partial_r \phi^i \partial_r \phi^j \Gamma_4 - g \partial_r (V_I h^I) \biggr] \epsilon = 0 \,.
\feq
We see immediately that one needs $\partial_r(V_I h^I)\neq0$, since otherwise
$\mathcal{G}_{ij}\partial_r\phi^i\partial_r\phi^j=0$, which is impossible because
$\mathcal{G}_{ij}$ is a definite matrix.

To proceed, we require the determinants associated to the linear systems \eqref{KSE:mag:rt} and 
\eqref{KSE:mag:rxyz} to vanish, which implies the projection condition
$\Gamma_4\epsilon=-\sigma_2\epsilon$ ($\sigma_2=\pm1$) as well as
\eq
\label{KSE:mag:rall}
\begin{gathered}
\sigma_2 g \partial_r (V_I h^I) \sqrt{V} = \frac12 V'' - g^2 (V_I h^I)^2 \,, \\
\sigma_2 g \partial_r (V_I h^I) \sqrt{V} = V T'' + V {T'}^2 \,.
\end{gathered}
\feq
Deriving the prefactor of $\epsilon$ in \eqref{KSE:mag:txyz} w.r.t.~$r$, one obtains, using also
\eqref{KSE:mag:rall}  and \eqref{KSE:mag:txyz},
\eq
\begin{split}
& 2g^2 (V_J h^J) \partial_r (V_I h^I) = \frac12 (V'' T' + V' T'') \\
& \qquad = \Bigl(\sigma_2 g \partial_r (V_I h^I) \sqrt{V} + g^2 (V_I h^I)^2\Bigr) T' + \frac12 V' \Bigl(\sigma_2 g \partial_r (V_I h^I) \frac{1}{\sqrt{V}} - {T'}^2\Bigr) \\
& \qquad = \sigma_2 g \partial_r (V_I h^I) \Bigl(\sqrt{V} T' + \frac{V'}{2\sqrt{V}}\Bigr)\,.
\end{split}
\feq
Thus, since $\partial_r(V_I h^I)\neq0$,
\eq
g(V_I h^I) = \sigma_2 \frac12 \Bigl(\sqrt{V} T' + \frac{V'}{2\sqrt{V}}\Bigr)\,.
\feq
Derive this w.r.t.~$r$ and then subtract the sum of the two eqns.~in \eqref{KSE:mag:rall}, divided by two,
to get
\eq
0 = \sigma_2 \frac12 \Bigl(\frac{V'T'}{\sqrt{V}} - \frac{{V'}^2}{4V^{3/2}} - \sqrt{V} {T'}^2\Bigr) = -\sigma_2 \frac{\sqrt{V}}{2} \Bigl(\frac{V'}{2V} - T'\Bigr)^2 = -\sigma_2 \frac{1}{2V^{3/2} {T'}^2} e^{-4T_3}\,,
\label{contradict-magn}
\feq
where the last step follows from the first eq.~of \eqref{T1=T2}. Evidently, \eqref{contradict-magn}
leads to a contradiction, which implies
\begin{proposition}
There are no static, Sol-invariant solutions to the Killing spinor equations with solvegeometry spatial
cross-sections at fixed $r$ and purely magnetic field strengths.
\end{proposition}
In particular, there is no BPS limit of the black hole constructed in section~\ref{sec:black-string}.
Note in this context that rotating supersymmetric Nil and $\widetilde{\text{SL}}(2,\bR)$ near-horizon 
geometries were found in \cite{Gutowski:2004ez}.
Moreover, it is worth mentioning that the near-horizon limit of all supersymmetric extremal black 
holes in gauged (and ungauged) five-dimensional supergravity coupled to abelian vector multiplets
must admit an $\text{SL}(2,\bR)$ symmetry group \cite{Kayani:2018yub}. This follows from an index
theory argument and extends earlier results of \cite{Grover:2013ima} for minimal gauged supergravity.

\section{Attractor mechanism}
\label{sec:attractor}

According to the attractor mechanism \cite{Ferrara:1995ih,Strominger:1996kf,Ferrara:1996dd,
Ferrara:1996um,Ferrara:1997tw}, the entropy of an extremal black hole and the scalar fields
on the event horizon are insensitive to the asymptotic values of the moduli and depend only on the
electric and magnetic charges. This phenomenon was first discovered in four-dimensional ungauged 
supergravity for BPS black holes \cite{Ferrara:1995ih} and subsequently extended to higher dimensions, 
non-supersymmetric or rotating solutions, and gauged supergravities, cf.~\cite{Goldstein:2005hq,
Astefanesei:2006dd,Morales:2006gm,Huebscher:2007hj,Bellucci:2008cb,Cacciatori:2009iz,
DallAgata:2010ejj,Kachru:2011ps,Chimento:2015rra} for an (incomplete) list of references.
A recurrent feature in all these cases is that the scalar configuration on the horizon can be determined
by extremizing an effective potential and that the entropy is given by the value of this potential at its
extremum.

In this section, we study the attractor mechanism for extremal static black holes with nil- or
solvegeometry horizons in the theory \eqref{lagrangian}.
It will turn out that there are no such attractors
for purely electric field strengths, while in the magnetic case there are attractor geometries, for which
we explicitely determine the effective potential $V_{\text{eff}}$, which contains the charges as well as
the scalar potential of the gauged supergravity theory.

\subsection{Magnetic ansatz}

As a first step to extend the black hole solution \eqref{blackstring} to the matter-coupled case,
we consider the near-horizon limit of the ansatz \eqref{ansatz-metric}.
Following closely the argument presented in \cite{Chimento:2015rra}, we are interested in magnetically
charged, static and extremal black holes with Sol horizon, but without referring to any particular
model of very special geometry.
Extremality implies that the near-horizon geometry is the product manifold
$\text{AdS}_2\times\text{Sol}$. Assuming the horizon to be located at $r=0$, we have thus for $r\to 0$
\eq
V(r)\sim\biggl(\frac r{r_\text{AdS}}\biggr)^2\,, \qquad T_1(r)\sim\frac14\ln A\,, \qquad T_3(r)\sim
\frac12\ln B\,, \qquad\phi^i(r)\sim\phi^i_0\,, \label{nearh}
\feq
with $r_\text{AdS}$ the curvature radius of the AdS$_2$ part, $A$ and $B$ positive constants and 
$\phi^i_0$ the horizon values of the scalar fields. The Einstein equations \eqref{einstein-magn} become
then algebraic and admit the solution
\eq
A = -\frac{\Sigma_0}{g^2 U_0}\,, \qquad  B = -\frac2{g^2 U_0}\,, \qquad r_\text{AdS}^2 =
-\frac1{g^2 U_0}\,, \label{attrsolution}
\feq
where $U_0\equiv U(\phi^i_0)<0$ and $\Sigma_0\equiv G_{IJ}(\phi^i_0)p^I p^J$. Using \eqref{nearh} 
and \eqref{attrsolution}, the equations \eqref{scalars-magn} for the scalars boil down to 
\eq
\label{extremization}
\partial_i V_{\text{eff}}\bigr|_{\phi^i_0} = 0\,,
\feq
where
\eq
\label{Veff}
V_{\text{eff}}(\phi^i) = \frac{\sqrt{G_{IJ}(\phi^i) p^I p^J}}{2\sqrt2 g^2 |U(\phi^i)|}
\feq
is an effective potential whose normalization has been chosen for later convenience. Thus, the
attractor solution reads
\eq
ds^2 = -g^2 |U_0| r^2 dt^2 + \frac{dr^2}{g^2 |U_0| r^2} + \sqrt{\frac{\Sigma_0}{g^2 |U_0|}}\left(e^{2z} 
dx^2 + e^{-2z} dy^2\right) + \frac2{g^2 |U_0|} dz^2\,,
\label{metricnh}
\feq
\eq
\label{vecfield}
F^I = p^I dx\wedge dy\,, \qquad \phi^i(r) = \phi^i_0\,.
\feq
The horizon values $\phi^i_0$ of the scalars are computed by extremization of the effective 
potential \eqref{Veff} and (unless $V_{\text{eff}}$ has flat directions) are completely fixed by the
magnetic charges and the constants $V_I$, in accordance with the attractor mechanism.
Finally, the entropy density is given by
\eq
\label{entropy-nh}
s = V_{\text{eff}}(\phi^i_0)\,.
\feq
Notice that, even if $V_{\text{eff}}$ has flat directions, and thus (some of) the moduli at the horizon
are not stabilized, \eqref{entropy-nh} implies that the
Bekenstein-Hawking entropy is given by the value of $V_{\text{eff}}$ at its minimum, which depends
only on the magnetic charges $p^I$ and the parameters $V_I$. As a consequence of the results of
section \ref{subsec:magBPS}, the attractor geometry \eqref{metricnh}, \eqref{vecfield} breaks all the
supersymmetries.

As an example, we consider the stu model, which involves two vector multiplets, and has
$C_{012}=1$ and its permutations as only nonvanishing components of $C_{IJK}$. We define
$t=\phi^1$, $u=\phi^2$, and choose the parametrization $h^1=t$, $h^2=u$ and $h^0=s=(tu)^{-1}$,
where the last relation follows from \eqref{hypersurface}.
Using the expresssions of section \ref{sec:gauged-sugra} and taking $V_I=1/3\;\forall I$, we get
\eq
G_{IJ} = \frac12\diag\bigl(s^{-2}, t^{-2}, u^{-2}\bigr) \,, \qquad
\mathcal{G}_{ij}=
\begin{bmatrix}
\dfrac{1}{t^2} & \; & \dfrac{1}{2tu} \\\\
\dfrac{1}{2tu} & \; & \dfrac{1}{u^2}
\end{bmatrix}\,,
\feq
\eq
U(t,u) = -2\biggl(tu + \frac{1}{t} + \frac{1}{u}\biggr)\,.
\feq
The effective potential \eqref{Veff} becomes
\eq
V_{\text{eff}}(t,u) = \frac{\sqrt{(p^0)^2 t^2 u^2 + (p^1)^2 t^{-2} + (p^2)^2 u^{-2}}}{8g^2\bigl(tu + t^{-1}
+ u^{-1}\bigr)}\,,
\label{Veff-stu}
\feq
and the eqns.~\eqref{extremization} boil down to
\begin{eqnarray}
(p^0)^2 t^3 u^3 (t+2u) - (p^1)^2 (u+2tu^3) - (p^2)^2 (t^3u-t)\bigr|_{t_0,u_0} &=& 0\,, \nonumber \\
(p^0)^2 t^3 u^3 (2t+u) - (p^1)^2 (tu^3-u) -(p^2)^2(t+2t^3u)\bigr|_{t_0,u_0} &=& 0\,.
\end{eqnarray}

\subsection{Electric ansatz}

We now consider the case of purely electric field strengths. For a horizon modelled on solvegeometry,
by means of the constraint \eqref{vincolo} and the structure constants \eqref{CabcSol}, the fourth
eq.~of \eqref{einstein-el} reduces to ($A=1,3$)
\eq
\begin{gathered}
-V' T_1' - V T_1'' - V T_1' (2T_1'+T_3') = \frac13 e^{-2(2T_1+T_3)} G^{IJ} q_I q_J + \frac23 g^2 U \,, \\
-V' T_3' - V T_3'' - V T_3' (2T_1'+T_3') - 2e^{-2T_3} = \frac13 e^{-2(2T_1+T_3)} G^{IJ} q_I q_J +
\frac23 g^2 U\,,
\label{eins1}
\end{gathered}
\feq
which immediately implies that a configuration with $T_1$ and $T_3$ constant is not
acceptable\footnote{For $T_1'=T_3'=0$, the difference of the two eqns.~leads to $e^{-2T_3} = 0$.}.
One can try to relax the ansatz on $T_1$ and $T_3$ by assuming a generic power dependence like
\eq
e^{2T_1} \sim k_1 r^{\alpha_1} \,, \qquad e^{2T_3} \sim k_3 r^{\alpha_3}\,,
\feq
with $k_A$ and $\alpha_A$ constants, but consistency of eqns.~\eqref{eins1} requires $\alpha_A=0$
and we fall into the previous contradictory case.

For a horizon modelled on nilgeometry, cf.~\eqref{CabcNil}-\eqref{Nilmanifold}, the fourth eq.~of 
\eqref{einstein-el} gives
\eq
\begin{gathered}
-V' T_1' - V T_1'' - V T_1'\sum_{B=1}^3 T_B' + \frac12 e^{2(T_1-T_2-T_3)} = \frac{q^2}3 
e^{-2(T_1+T_2+T_3)} + \frac23 g^2 U \,, \\
-V' T_2' - V T_2'' - V T_2'\sum_{B=1}^3 T_B' - \frac12 e^{2(T_1-T_2-T_3)} = \frac{q^2}3 
e^{-2(T_1+T_2+T_3)} + \frac23 g^2 U \,, \\
-V' T_3' - V T_3'' - V T_3'\sum_{B=1}^3 T_B' - \frac12 e^{2(T_1-T_2-T_3)} = \frac{q^2}3 
e^{-2(T_1+T_2+T_3)} + \frac23 g^2 U \,,
\label{eins2}
\end{gathered}
\feq
where $q^2=G^{IJ}q_Iq_J$. Again, an ansatz with $T_1$, $T_2$ and $T_3$ constant does not 
work, since in that case the difference of the first and the second eq.~of~\eqref{eins2} yields
\eq
e^{2(T_1-T_2-T_3)} = 0\,.
\feq
If we assume $e^{2T_A}\sim k_A r^{\alpha_A}$ and plug this ansatz into \eqref{eins2}, we end up with $\alpha_A=0$, which we have just seen to lead to a contradiction. One obtains thus the following
\begin{proposition}
There are no static attractors with Sol or Nil horizons and purely electric field strengths.
\end{proposition}

\section*{Acknowledgements}
This work was supported partly by INFN and by MIUR-PRIN contract 2017CC72MK003.

\appendix

\section{Homogeneous manifolds}
\label{app:homogeneous-spaces}

Let $M$ be a (pseudo)-Riemannian manifold with isometry group $G$. $M$ is said to be homogeneous
if $G$ acts transitively on $M$, i.e.~if $\forall\,p,q\in M$ there exists an isometry $\phi \in G$ such that $\phi(p)=q$. The action of $G$ on $M$ is called simply transitive if the element $\phi$ is unique or, equivalently, if $\dim M=\dim G$. In this case, $M$ itself is said to be simply transitive.

Let us restrict our discussion to a simply transitive manifold. Since $\dim M=\dim G$, the Killing vectors $\xi_A$ ($A=1,\ldots,\dim M$) form a basis of the tangent space. However, it is more
convenient \cite{Ryan:1975jw} to choose a $G$-invariant basis $X_A$, i.e., a basis such that
\eq
\mathcal{L}_{\xi_B} X_A = [\xi_B, X_A] = 0\qquad\forall\,A,B\,,  
\label{G-inv}
\feq
with $\mathcal{L}_{\xi_B}X_A$ the Lie derivative of the vector field $X_A$ along $\xi_B$. The dual basis 
$\theta^A$ of a $G$-invariant basis $X_A$ is also $G$-invariant, $\mathcal{L}_{\xi_B}\theta^A=0$,
and satisfies
\eq
d\theta^A = \frac12 C^A_{\ BC} \theta^B\wedge\theta^C\,,
\feq
with $C^A_{\ BC}$ the structure constants of the Lie algebra of $G$. Furthermore, a simply transitive 
homogeneous manifold can be equipped with a metric
\eq
ds^2 = g_{AB}\theta^A\theta^B\,,
\feq
where the components $g_{AB}$ are constant on $M$.

Bianchi showed that in total there are nine three-dimensional Lie algebras, the so-called nine Bianchi 
cosmologies, labelled from type I to type IX. The name `cosmologies' comes from the fact that these
manifolds are used as spatial sections in many spatially homogeneous but anisotropic cosmological
models. The Bianchi cosmologies are divided into two classes, A and B, according to the way the structure 
constants $C^A_{\ BC}$ can be expanded (see table 6.2 of \cite{Ryan:1975jw} for details). In particular,
class A spacetimes satisfy
\eq
\label{classA}
\sum_A C^A_{\ AB} = 0\,.
\feq
An important result in geometric topology is the Thurston conjecture \cite{Thurston:1997}, which states
that every three-dimensional closed and orientable manifold has a geometric structure modelled on one
of the eight model geometries
\eq
\text{S}^3\,, \quad\text{E}^3\,, \quad\text{H}^3\,, \quad\text{S}^2\times\mathbb{R}\,, \quad 
\text{H}^2\times\mathbb{R}\,, \quad\text{Nil}\,, \quad\text{Sol}\,, \quad\widetilde{\text{SL}}
(2,\mathbb{R})\,,
\feq
where $\widetilde{\text{SL}}(2,\mathbb{R})$ is the universal covering of $\text{SL}(2,\mathbb{R})$.
In \cite{Fagundes:1991uy} it was shown that there exists a correspondence, not necessarily one to one, 
between the nine Bianchi cosmologies and the eight Thurston model geometries, which is summarized in
table \ref{table:bianchi-thurston}\footnote{The Bianchi types IV and VI$_{h\ne-1}$ are not contained
in this correspondence. Moreover, the Thurston geometry $\text{S}^2\times\mathbb{R}$ is missing
since it corresponds to the Kantowski-Sachs model, in which $G$ does not act simply transitively or does
not possess a subgroup with simply transitive action.}.
\begin{table}[H]
\begin{center}
\begin{tabular}[t]{| p{5em} | p{5em} |}
  \hline
  Bianchi	&Thurston \\
  \hline\hline
  I, VII$_{0}$	& $\text{E}^3$  \\ \hline
  II		& Nil \\ \hline
  VI$_{-1}$	& Sol \\ \hline
  VIII		& $\widetilde{\text{SL}}(2,\mathbb{R})$ \\ \hline
  IX		& $\text{S}^3$ \\  
  \hline
\end{tabular}
\qquad
\begin{tabular}[t]{| p{5em} | p{5em} |}
  \hline
  Bianchi		& Thurston \\
  \hline \hline
  III			& $\text{H}^2\times\mathbb{R}$ \\ \hline
  V, VII$_{h\ne 0}$	& $\text{H}^3$ \\
  \hline
\end{tabular}
\end{center}
\caption{Class A (left) and B (right) spacetimes and corresponding Thurston geometries.}
\label{table:bianchi-thurston}
\end{table}
In the following, we list explicitely the metrics for solvegeometry/VI$_{-1}$ and nilgeometry/II in terms
of $G$-invariant one-forms $\theta^A$, as well as the nonvanishing structure constants of the related
Lie algebras.
\begin{itemize}
\item{Solvegeometry:}
\eq
\label{CabcSol}
C^1_{\ 13} = -C^1_{\ 31} = 1\,, \qquad C^2_{\ 23} = -C^2_{\ 32} = -1\,,
\feq
\eq
\label{Solforms}
\theta^1 = e^z dx\,, \qquad\theta^2 = e^{-z} dy\,, \qquad\theta^3 = -dz\,,
\feq
\eq
\label{Solmanifold}
ds^2 = e^{2z} dx^2 + e^{-2z} dy^2 + dz^2\,.
\feq
\item{Nilgeometry:}
\eq
\label{CabcNil}
C^1_{\ 23} = -C^1_{\ 32} = 1\,,
\feq
\eq
\theta^1 = dz - x dy\,, \qquad\theta^2 = dy\,, \qquad\theta^3 = dx\,,
\feq
\eq
\label{Nilmanifold}
ds^2 = (dz - xdy)^2 + dy^2 + dx^2\,.
\feq
\end{itemize}

\end{document}